\documentclass[10pt]{article}

\usepackage{amsmath}
\usepackage{amssymb}

\usepackage{graphicx}

\usepackage{cite}

\usepackage{color} 

\usepackage[labelfont=bf,labelsep=period,justification=raggedright]{caption}

\makeatletter
\renewcommand{\@biblabel}[1]{\quad#1.}
\makeatother

\date{}

\definecolor{pink}{rgb}{1,1,0} 
\definecolor{red}{rgb}{1,0,0}
\definecolor{yellow}{rgb}{1,1,0}
\definecolor{orange}{rgb}{1,0.5,0} \definecolor{white}{rgb}{1,1,1}
\definecolor{blue}{rgb}{0,0,1}

\begin{document}

\begin{flushleft}
{\Large
\textbf{Mapping the Evolution of Scientific Fields}
}
\\
Mark Herrera$^{1}$, 
David C. Roberts$^{2,\ast}$, 
Natali Gulbahce $^{3,\ast}$
\\
\bf{1} Department of Physics and Institute for Research in Electronics and Applied Physics, University of 
Maryland, College Park, MD, USA
\\
\bf{2} Theoretical Division and Center for Nonlinear Studies, Los Alamos 
National Laboratory, Los Alamos, USA

\bf{3} Department of Physics and Center for
Complex Networks Research, Northeastern University, Boston, MA, USA 
\\
Center for Cancer Systems Biology, Dana Farber Cancer Institute, Boston, MA \\
$\ast$ E-mail: dcr@lanl.gov, natali.gulbahce@gmail.com
\end{flushleft}

\section*{Abstract}
Despite the apparent cross-disciplinary interactions among scientific fields, a
formal description of their evolution is lacking. Here we describe a novel
approach to study the dynamics and evolution of scientific fields using a
network-based analysis. We build an {\it idea} network consisting of American
Physical Society Physics and Astronomy Classification Scheme (PACS) numbers as
nodes representing scientific concepts. Two PACS numbers are linked if there
exist publications that reference them simultaneously. We locate scientific
fields using a community finding algorithm, and describe
the time evolution of these fields over the course of 1985-2006. The communities we identify map
to known scientific fields, and their age depends on their size and
activity. We expect our approach to quantifying the evolution of ideas to be
relevant for making predictions about the future of science and thus help to
guide its development. 

\section*{Introduction}

Cross-fertilization between different scientific fields has been recognized for
its ability to encourage new developments and innovative thinking. For this
reason, multidisciplinary approaches to research are becoming more popular.
Some recent examples include applying physics techniques to the study of
biological phenomena \cite{frauenfelder}, deriving an understanding of the
nature of critical phenomena from renormalization techniques in particle
physics \cite{wilson} drawing inferences about the early universe from findings
in terrestrial superfluid experiments \cite{zurek}, and using statistical
physics to analyze technological and social systems \cite{dorogovtsev}. 

In an effort to move beyond anecdotal evidence of the benefit of
interdisciplinary discourse for science, in this paper we study the dynamics of
groups, or ``communities", of ideas using a statistical physics approach.  We
attempt to quantify the evolution of ideas and subdisciplines within physics as
they emerge, interact, merge, stagnate, and desist. The quest for describing
the development of scientific fields is not new. There have been
epidemiological \cite{goffman, tabah} and network-based approaches (citation
and collaboration networks)
~\cite{price,newmancoauth1,newmancoauth2,newmancoauth3,lehmann,borner,leydesdorff,bollen1,boyack} aiming
to gain insight into the spread of scientific ideas. Recently the temporal
evolution of several scientific disciplines have been modeled with a
coarse-grained approach \cite{luis}. 

Here we build a scientific concept network consisting of American Physical
Society PACS numbers as nodes representing scientific concepts.   The American Institute of Physics (AIP) develops and maintains the PACS scheme as a service to the physics community in aiding the
classification of scientific literature and information retrieval.  Two PACS numbers are linked if there
exist publications that reference them simultaneously. Our approach differs
from previous methods in that it provides a direct, unsupervised description of
scientific fields and uses techniques such as community finding and tracking
from the field of network physics. This approach provides means to quantify how
ideas and movements in science appear and fade away.  Because this method makes
it possible to measure the current and past state of the relationship between
scientific concepts, it may also help to make predictions about the future of
science and thus inform efforts to guide its development. In this paper, we
entertain some of the quantitative questions that this method permits;
specifically, we seek to answer questions about the relationship between size,
lifetime, and activity of scientific fields.

Various local to global topological measures have been introduced to unveil the
organizational principles of complex networks
\cite{albertrev,newmanbook,caldarelli}. One such measure that allows the
discovery of organizational principles of networks is community finding.  There
have been a number of methods to find the communities in networks which
describe the inherent structure or functional units of a network
\cite{newmangirvan, palla1, clauset, gulbahce}. One of these is CFinder, a
clique percolation method (CPM) introduced by Palla et al. \cite{palla1}, which finds
overlapping communities and is especially suitable for studying the evolution
of scientific fields since scientific concepts are often shared among multiple
fields. We use this CPM to track the evolution of physics.

\section*{Results}

\noindent{\bf Building the Network}

Data were collected from the American Physical Society's (APS) {\it Physical
Review} database from 1977-2007. Journals included in the study are {\it
Physical Review Letters}, {\it Physical Review \{A} through {\it E\}}, and {\it
Physical Review Special Topics: Accelerators and Beams}. Papers in this
database contain a list of author-assigned PACS codes, where each PACS code
refers to a specific topic in physics.  PACS itself is hierarchical, which is
evident in the structure of the codes with up to 5 levels of topic
specification.  For example the PACS code `64.60.aq' has 5 levels where the
first digit `6' represents the first level (in this case `condensed matter'),
`4' represents the second (e.g. `equations of state, phase equilibria, and
phase transitions'), the third and fourth digits `60' together represent the
third level (e.g., `general studies of phase transitions') while the last two
characters `aq' carry information pertaining to the fourth and fifth levels of
specification (e.g. `specific approaches applied to phase transitions' and
`networks', respectively).

PACS codes are not static, rather,  the coding scheme is periodically updated with the addition and deletion of codes.  In order to (at least partially) account for this effect, the scientific concept network was constructed such that the nodes in the network represent individual PACS codes using the first four
digits of specification, where changes to scheme are less probable. This network and the related material is available on
our website \cite{datalink}.  In our network, an {\it edge} occurs between two
nodes if the two PACS codes they represent are cited in the same paper; one
paper in the database often contributes many nodes and edges to the
network. Furthermore, edges are weighted by the number of papers that contain
that edge. We introduce two measures, node and edge cutoffs, to control for noise 
in the network (see Methods section).

\begin{figure}[h] 
\begin{center}
\scalebox{1}{\includegraphics{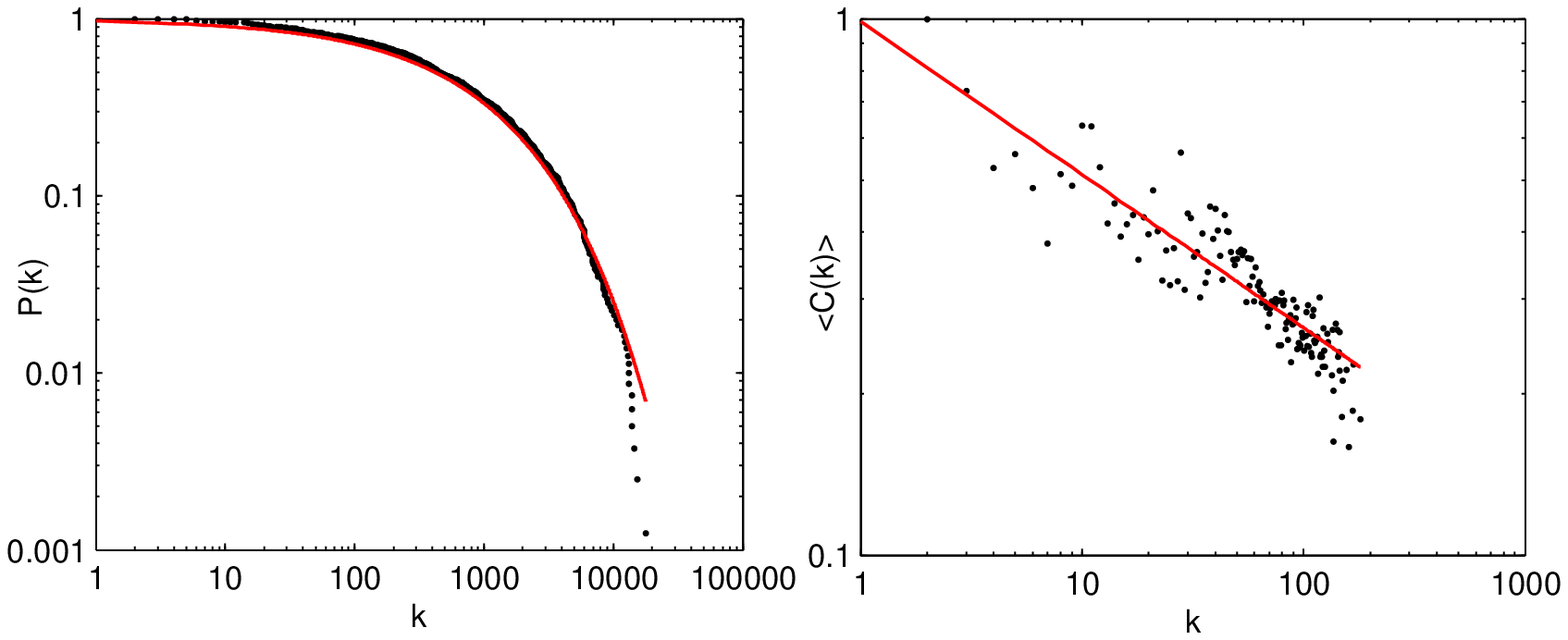}}
\caption{ {\bf Measurements on the PACS network from 1977-2007.}  A) Cumulative degree distribution P(k) of the PACS
network. The red line is a fit to the data. Both the weighted and unweighted cases 
follow a stretched exponential distribution.  {B)\bf Average clustering vs degree 
for the PACS network, demonstrating that $C(k)$ has some dependence on degree.} Thus, 
there is some hierarchical structure present in the network.}
\label {DegreeFig}
\end{center}
\end{figure}

The entire PACS network from 1977-2007 after both noise measurements were
applied has 803 nodes and 23707 distinct edges.  The \emph{degree} of a node is
the number of edges shared by the node.  The weighted cumulative degree
distribution follows a stretched exponential with the form,  $P(k)\sim\
\mbox{exp} [-(k/842)^{0.53}]$ as shown in Fig. \ref{DegreeFig}A. The
distribution has a similar form in the unweighted case.  The dynamic
classification scheme of the American Physical Society, implemented by the
addition, splitting and removal of codes, may be preventing the formation of
large hubs, thus keeping the specification of the codes more useful. The
stretched exponential distribution may be the result of a sublinear-linear
attachment type growth \cite{krapivsky}.

The PACS network also exhibits a weak but apparent hierarchical structure
measured by the dependence of the \emph{clustering coefficient} on (unweighted)
degree.  For a node $i$, the clustering coefficient is given as $C_i =
2n_i/k_i(k_i-1)$, where $n_i$ is the number of edges that link the neighbors of
node $i$, and $k_i$ is the degree of the node. The clustering coefficient for a
node is the ratio of the number of triangles through node $i$ over the possible
number of triangles that could pass through node $i$ \cite{barabasi}.  A purely
hierarchical network will have a $\langle C \rangle$ that scales as a power of
$k$, $\langle C \rangle \sim k^{-1}$, while a random network will have a
clustering coefficient that is constant with $k$ \cite{barabasi}.  For this
network,  $\langle C(k) \rangle \sim k^{-0.29}$ , shown in Fig. 
\ref{DegreeFig}B. This dependence is not surprising given the hierarchical
structure of the classification scheme.

\noindent{\bf Defining Communities in Physics}

Papers published between 1985 and 2006 were used to study the community
evolution of the network; 1985 appears to be the first year when all journals present ({\it Physical Review E} began publication in 1993)
consistently used the PACS data scheme, and 2007 was thrown out to exclude
incomplete data from the analysis. The journal {\it Physical Review Special
Topics: Accelerators and Beams} was not included because of an irregular
publishing schedule. After the noise measures were carried out, the edge
weights were no longer used, and the network became an unweighted network with
respect to the community evolution analysis. The data were organized into 44
time bins, with each bin representing a 0.5 year time period. Once a paper (and
the edges and nodes it contains) appears in the analysis, it is assigned a
lifetime, $l$, of 0 or 2.5 years. This assignment is an attempt to more
realistically capture the nature of scientific dissemination, as well as the
delay in time from publication to assimilation by the field. The analysis of
community evolution begins at the time bin subsequent to the lapse of the
assigned lifetime.  Thus the first time bin, $t=0$, for a paper lifetime of
$l=2.5$ refers to the latter half of 1987 since we start the analysis in 1985.  

\begin{figure}[ht] 
\begin{center} 
\scalebox{1}{\includegraphics{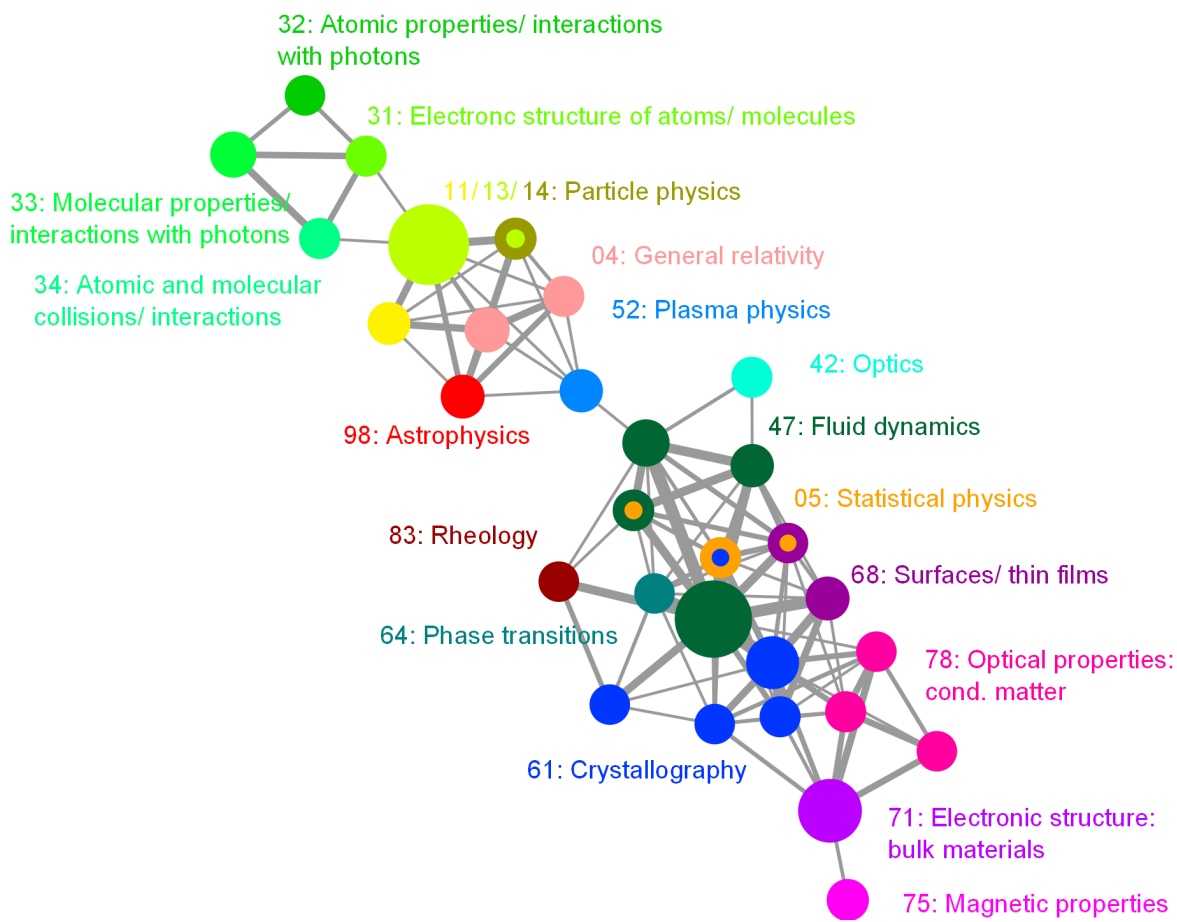}}
\caption{ {\bf The scientific concept network for the first half of 1997}. Nodes corresponding to
scientific fields, as well as node labels and their corresponding fields, are shown in
the same color. The size of the nodes corresponds to the number of PACS codes contained in that community.  Same-color neighboring nodes have the same label. The thickness of the edges correspond to the number of shared PACS codes between communities (the weight of the edge).  The community structure is shown at $t=9.5$ years, corresponding to first half of 1997, using CFinder with  $l=2.5$ years. Labels
are assigned by looking at the first two digits of the PACS codes that make up
the largest fraction of each community. } 
\label {comdiag}
\end{center} 
\end{figure}

\begin{figure}[ht] 
\begin{center} 
\scalebox{1}{\includegraphics{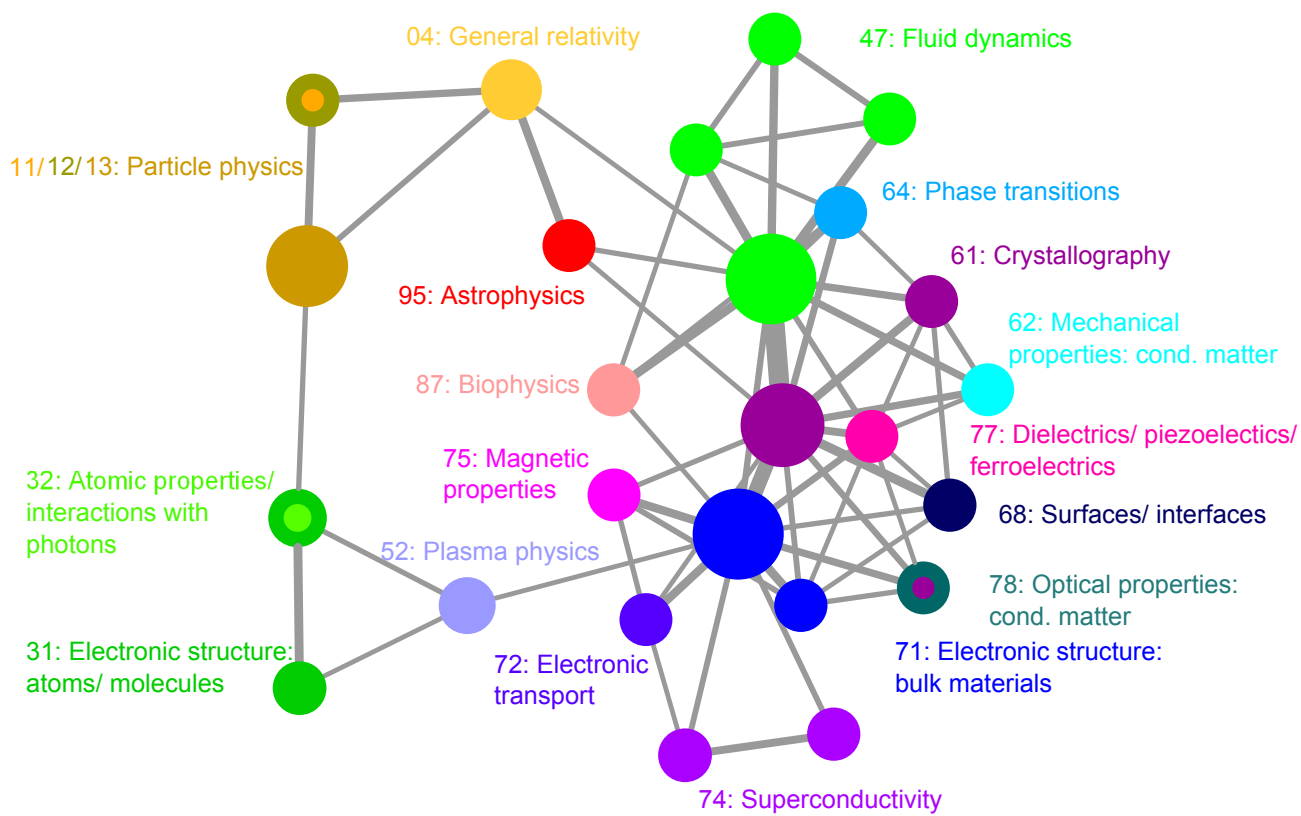}}
\caption{{\bf The scientific concept network for the first half of 2005.}}
\label {comdiaga} 
\end{center} 
\end{figure}

In order to study the evolution of different fields in physics, one must first
find these fields in our network. We hypothesize that scientific fields are
represented by communities in our PACS network. These communities are found
using the CFinder algorithm, which is based on a clique percolation
method\cite{palla1}.  Figs. \ref{comdiag} and \ref{comdiaga}  present
examples of the community structure extracted utilizing CFinder.  

For each community, the code (using only the first two digits) that encompasses
the largest fraction of nodes in the community was found. Its name, specified by
the PACS scheme, is then used to label the community.  If a community has
multiple codes which compose the same largest fraction of nodes in that
community, then the community is assigned multiple labels. As shown in
Figs.~\ref{comdiag} and ~\ref{comdiaga}, we observe that the analysis captures 
expected scientific connections among fields in physics.  For example, in 1997, particle physics is linked to both general relativity and astrophysics.  It is also worthwhile to note the emergence of biophysics as a community in the 2005 analysis.  \\ 

\noindent{\bf Community Evolution and Dynamics}

In order to track the evolution of scientific fields, after identifying
communities at each individual time interval, it is necessary to match the
communities between adjacent time steps. We implemented a community evolution
algorithm developed by Palla et al.~\cite{palla2} to match the communities
between time bins (see Methods section).

To gain a better understanding of the dynamics of evolving communities, we
defined two properties of each community: size and activity. A value for each of
these measures can be assigned to every community for each individual time bin.
The size $s$ of a community is the number of nodes contained within that
community at time $t$. Size can be interpreted as a measure of a community's
breadth: communities with a small size encompass only a few distinct ideas,
while large communities encompass many distinct ideas.  (The cumulative size distribution was calculated for different times and is displayed in Fig. S1.)

The activity $\alpha$ of a community is defined as the number of papers that
contain at least one node from that community at time $t$. As one expects,
there is a strong correlation between size and activity (see Fig. S2).

\begin{figure}[h!]
\begin{center}
\scalebox{1.0}{\includegraphics{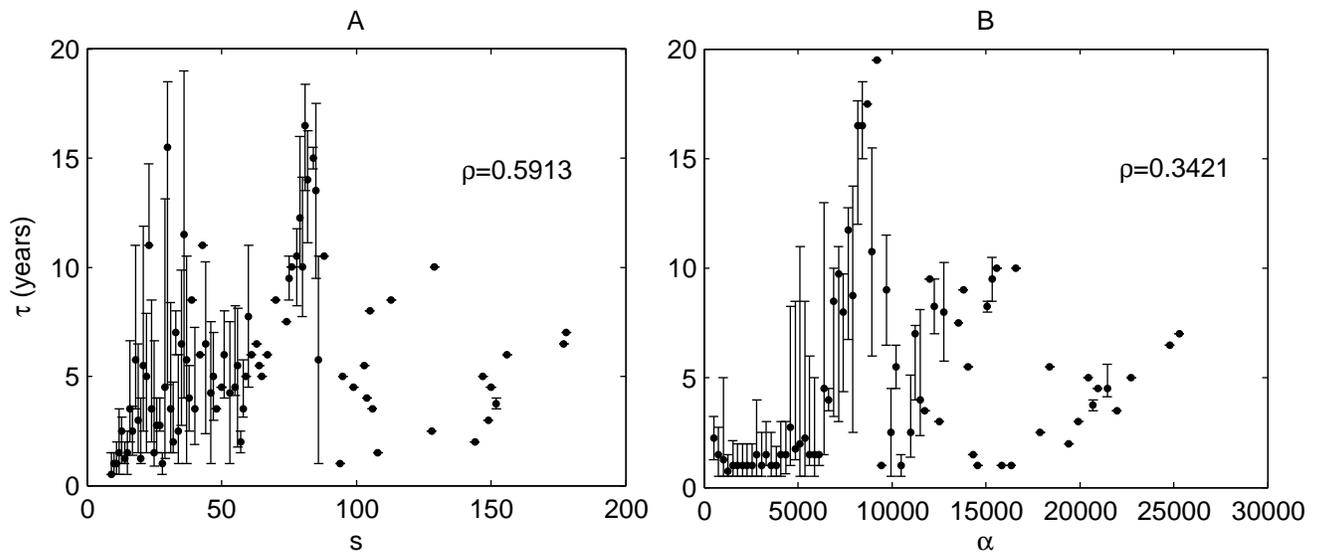}}
\caption{
{\bf For $l=2.5$ years, the median lifetime (years) as a function of A) size;  B)
activity $(\alpha)$}.  Error bars represent the 1st and 3rd quartiles respectively.   For both sets of data, the Spearman's rank correlation coefficient, $\rho$, was computed using the unbinned data.}
\label{fig2} 
\end{center} 
\end{figure}

Next, we study the relationship between the age or lifetime of a community
versus its size and activity. The age of a community at time $t$ is simply the
number of time bins the community has been present in the evolution analysis:
$\tau=t-t_0+1$, where $t_0$ is the time bin in which the community was born. In
order to study the dependence of age on size, in each time bin, the current age
$\tau$ and size $s$ are recorded. Using all communities from all time
intervals, the median age is calculated for communities with the same size as
shown in Fig.  \ref{fig2}A.  There is a trend of $\tau$ increasing with
size $s$.  Thus, it would appear that older communities tend to contain more
nodes, and that longer lived fields tend to encompass many distinct ideas.   Values for both the Pearson correlation coefficient, $p$, and the Spearman's rank correlation coefficient, $\rho$, were calculated between $\tau$ and $s$ using the raw, unbinned data.  $\rho=1-6\sum_i\frac{d_i^2}{N(N^2-1)}$ where $N$ is the number of data points and $d_i$ is the difference in the statistical rank of the corresponding values for each data point.  For $l=2.5$, the Pearson correlation coefficient was $p=0.4772$ while the  Spearman's $\rho$ was calculated to be $ \rho=0.5913$.

In order to measure the dependence of age on activity, the current age $\tau$
is recorded along with the current activity $\alpha$ of every community in each
time step. Because of the wide range of possible values for activity and noise
in the data, the values of $\alpha$  are sorted into 100 equally sized bins.
The median age is calculated for all communities within the same activity
interval. There is a trend of $\tau$ increasing with activity as shown in
Fig.  \ref{fig2}B which can be partially understood by the strong correlation
between size and activity. Further we note an apparent phase transition in
activity; as shown in Fig.  \ref{fig2}B after some critical value, communities
tend to be longer lived.   This transition also appears for $l=0$ (see Fig. S3).  Lifetime as a function of size, $\tau(s)$, for $l=0$ is shown in Fig. S4. Again, the Pearson correlation coefficient and the Spearman's rank correlation coefficient were calculated for $l=2.5$ using the raw, unbinned data between $\tau$ and $\alpha$, with $p=0.3283$ and $\rho=0.3764$.

\section*{Discussion}

In this paper, we have developed an approach that enables the quantitative
study of the evolution of physics fields, specifically by following the
dynamical connections between various ideas within physics. From our
investigation, we have shown that long lived communities tend to be larger, and
are associated with a higher number of papers. 

Our approach opens up an interesting possibility of being able to predict
community dynamics and impact from the current network structure. Furthermore,
this method can be easily adapted to other scientific fields using different
databases. One such is the INSPEC database which has comprehensive coverage of
research activity in computer science and engineering in addition to physics,
and has an expert-assigned classification scheme rather than author-based
assignments. 


\section*{Materials and Methods}
\subsection*{Noise Measures }
A node cutoff is introduced such that in a given time interval a node must
appear at least twice to be included in the network. This measure eliminates
many of the typographical errors occurring in the database. The edge cutoff,
however, takes into account the random expectation of two PACS codes
co-occurring in the same paper. For this cutoff, the weight of an edge between
nodes $i$ and $j$, $W_{ij}$, which is the number of papers that both codes $i$
and $j$ appear in, is compared to the weight expected at random, $E_{ij}=n_i
n_j/N$, where $n_i$ and $n_j$ are the number of papers containing nodes $i$ and
$j$ respectively, and $N$ is the total number of papers present in the time
interval.   If $W_{ij}/E_{ij} > 1.2$, then the appearance of the edge is
significant compared to random appearance, and we include it in the network.\\ 
\subsection*{CFinder} 
The CFinder algorithm is described in detail in Ref.
\cite{palla1}.  A community is defined as a union of all $k$-cliques (complete
subgraphs of size $k$) that can be reached from each other through a series of
adjacent $k$-cliques (where adjacency means sharing $k - 1$ nodes)
\cite{palla1}.  \\ 
\subsection*{Picking a $k$ value}
For this study, $k=9$ was principally used (for l=2.5) because it appears to produce a
large number of communities while discouraging the formation of giant
communities. Further, by keeping $k$ constant, we keep the resolution constant
for the entire analysis. Picking an appropriate $k$ value for the analysis is
done by considering two properties: the number of communities present, and the
presence of overly large communities \cite{palla1}. It is desirable to have a
large number of communities, so as to increase the statistical quality of
measurements made on the network. Fig.  S6 plots the number of
present communities for each time step for  $k=8, 9$, and $10$, for $l=2.5$. As
demonstrated, the number of communities found using the choice of $k=10$ tends
to be less than the other parameter choices, making it less favorable in terms
of improving statistical quality.

A $k$ value must also be large enough to avoid the introduction of overly large
communities that obscure the actual community structure of the network
\cite{palla1}. To quantify this property, we use the quantity $r$ which is the
ratio of the size of the largest community to the second largest community for a
given time bin.  Thus while some distribution in the sizes of communities is
necessary, $r$ should not be overly large.   Fig.  S7 plots the measure
$r$ against all time bins for $l=2.5$.  For $k=8$, the values of $r$ tends to be larger than (signifying giant communities) than those calculated from the other two parameter values, making it an unfavorable parameter choice. 
 
\subsection*{Community Matching}
The community matching algorithm is described in detail in Ref. \cite{palla2}.
In this analysis, an appropriate $k$-value is used rather than a constant
edge-weight cutoff.  A running stationarity measure is described in Appendix S1 and Figure S5.  The merger of two communities is described in Appendix S1 and Figures S8 and S9.
\section*{Acknowledgments} 
The authors would like to thank the American Physical Society and the American
Institute of Physics for the use of their data. The authors would like to
acknowledge Gergely Palla, Sune Lehmann, Albert-L\'aszl\'o Barab\'asi, Tam\'as
Vicsek, Aric Hagberg, Hristo Djidjev, Luis Bettencourt, and Michael Ham for
useful discussions. 

\bibliography{plos_new}

\begin{thebibliography}{10}
\providecommand{\url}[1]{\texttt{#1}}
\providecommand{\urlprefix}{URL }
\expandafter\ifx\csname urlstyle\endcsname\relax
  \providecommand{\doi}[1]{doi:\discretionary{}{}{}#1}\else
  \providecommand{\doi}{doi:\discretionary{}{}{}\begingroup
  \urlstyle{rm}\Url}\fi
\providecommand{\bibAnnoteFile}[1]{%
  \IfFileExists{#1}{\begin{quotation}\noindent\textsc{Key:} #1\\
  \textsc{Annotation:}\ \input{#1}\end{quotation}}{}}
\providecommand{\bibAnnote}[2]{%
  \begin{quotation}\noindent\textsc{Key:} #1\\
  \textsc{Annotation:}\ #2\end{quotation}}
\providecommand{\eprint}[2][]{\url{#2}}

\bibitem{frauenfelder}
Frauenfelder H, Wolynes PG, Austin RH (1999) Biological physics.
\newblock Rev Mod Phys 71: S419--S430.
\bibAnnoteFile{frauenfelder}

\bibitem{wilson}
Wilson KG (1975) The renormalization group: Critical phenomena and the kondo
  problem.
\newblock Rev Mod Phys 47: 773--840.
\bibAnnoteFile{wilson}

\bibitem{zurek}
Zurek WH (1985) Cosmological experiments in superfluid helium.
\newblock Nature 317: 505-508.
\bibAnnoteFile{zurek}

\bibitem{dorogovtsev}
Dorogovtsev SN, Goltsev AV, Mendes JFF (2008) Critical phenomena in complex
  networks.
\newblock Reviews of Modern Physics 80: 1275.
\bibAnnoteFile{dorogovtsev}

\bibitem{goffman}
Goffman W, Harmon G (1971) Mathematical approach to the prediction of
  scientific discovery.
\newblock Nature 229: 103--104.
\bibAnnoteFile{goffman}

\bibitem{tabah}
Tabah AN (1999) Literature dynamics: Studies on growth, diffusion, and
  epidemics.
\newblock Annual Review of Information Science and Technology 34: 249-286.
\bibAnnoteFile{tabah}

\bibitem{price}
de~Solla~Price DJ (1965) Networks of scientific papers.
\newblock Science 149: 510-515.
\bibAnnoteFile{price}

\bibitem{newmancoauth1}
Newman MEJ (2001) {The structure of scientific collaboration networks}.
\newblock Proceedings of the National Academy of Sciences of the United States
  of America 98: 404-409.
\bibAnnoteFile{newmancoauth1}

\bibitem{newmancoauth2}
Newman MEJ (2001) Scientific collaboration networks. i. network construction
  and fundamental results.
\newblock Phys Rev E 64: 016131.
\bibAnnoteFile{newmancoauth2}

\bibitem{newmancoauth3}
Newman MEJ (2001) Scientific collaboration networks. ii. shortest paths,
  weighted networks, and centrality.
\newblock Phys Rev E 64: 016132.
\bibAnnoteFile{newmancoauth3}

\bibitem{lehmann}
Lehmann S, Lautrup B, Jackson AD (2003) Citation networks in high energy
  physics.
\newblock Phys Rev E 68: 026113.
\bibAnnoteFile{lehmann}

\bibitem{borner}
HerrII BW, Duhon RJ, B\"orner K, Hardy EF, Penumarthy S (2008) 113 years of
  physical review: Using flow maps to show temporal and topical citation
  patterns.
\newblock International Conference on Information Visualisation : 421-426.
\bibAnnoteFile{borner}

\bibitem{leydesdorff}
Leydesdorff L (2007) Betweenness centrality as an indicator of the
  interdisciplinarity of scientific journals.
\newblock J Am Soc Inf Sci Technol 58: 1303--1319.
\bibAnnoteFile{leydesdorff}

\bibitem{bollen1}
Bollen J, Van~de Sompel H, Hagberg A, Bettencourt L, Chute R, et~al. (2009)
  Clickstream data yields high-resolution maps of science.
\newblock PLoS ONE 4: e4803.
\bibAnnoteFile{bollen1}

\bibitem{boyack}
Boyack KW, Klavans AR, B{\"o}rner BK (2005) Mapping the backbone of science.
\newblock Scientometrics 64: 351--374.
\bibAnnoteFile{boyack}

\bibitem{luis}
Bettencourt L, Kaiser D, Kaur J, Castillo-Ch\'{a}vez C, Wojick D (2008)
  Population modeling of the emergence and development of scientific fields.
\newblock Scientometrics 75: 495--518.
\bibAnnoteFile{luis}

\bibitem{albertrev}
Albert R, Barab\'asi AL (2002) Statistical mechanics of complex networks.
\newblock Rev Mod Phys 74: 47--97.
\bibAnnoteFile{albertrev}

\bibitem{newmanbook}
Newman M, Barabasi AL, Watts DJ (2006) The Structure and Dynamics of Networks:
  (Princeton Studies in Complexity).
\newblock Princeton, NJ, USA: Princeton University Press.
\bibAnnoteFile{newmanbook}

\bibitem{caldarelli}
Caldarelli G (2007) Scale-free networks: complex webs in nature and technology.
\newblock Oxford University Press.
\bibAnnoteFile{caldarelli}

\bibitem{newmangirvan}
Girvan M, Newman MEJ (2002) {Community structure in social and biological
  networks}.
\newblock Proceedings of the National Academy of Sciences of the United States
  of America 99: 7821-7826.
\bibAnnoteFile{newmangirvan}

\bibitem{palla1}
Palla G, Derenyi I, Farkas I, Vicsek T (2005) Uncovering the overlapping
  community structure of complex networks in nature and society.
\newblock Nature 435: 814--818.
\bibAnnoteFile{palla1}

\bibitem{clauset}
Clauset A, Moore C (2008) Hierarchical structure and the prediction of missing
  links in networks.
\newblock Nature 453: 98-101.
\bibAnnoteFile{clauset}

\bibitem{gulbahce}
Gulbahce N, Lehmann S (2008) The art of community detection.
\newblock Bioessays 30: 934-938.
\bibAnnoteFile{gulbahce}

\bibitem{datalink}
http://nuweb6neuedu/ngulbahce/pacsdatahtml.
\bibAnnoteFile{datalink}

\bibitem{krapivsky}
Krapivsky PL, Redner S, Leyvraz F (2000) Connectivity of growing random
  networks.
\newblock Phys Rev Lett 85: 4629--4632.
\bibAnnoteFile{krapivsky}

\bibitem{barabasi}
Barabasi AL, Oltvai ZN (2004) Network biology: understanding the cell's
  functional organization.
\newblock Nat Rev Genet 5: 101--113.
\bibAnnoteFile{barabasi}

\bibitem{palla2}
Palla G, Barabasi AL, Vicsek T (2007) Quantifying social group evolution.
\newblock Nature 446: 664--667.
\bibAnnoteFile{palla2}

\end{thebibliography}


\begin{thebibliography}{1}
\providecommand{\url}[1]{\texttt{#1}}
\providecommand{\urlprefix}{URL }
\expandafter\ifx\csname urlstyle\endcsname\relax
  \providecommand{\doi}[1]{doi:\discretionary{}{}{}#1}\else
  \providecommand{\doi}{doi:\discretionary{}{}{}\begingroup
  \urlstyle{rm}\Url}\fi
\providecommand{\bibAnnoteFile}[1]{%
  \IfFileExists{#1}{\begin{quotation}\noindent\textsc{Key:} #1\\
  \textsc{Annotation:}\ \input{#1}\end{quotation}}{}}
\providecommand{\bibAnnote}[2]{%
  \begin{quotation}\noindent\textsc{Key:} #1\\
  \textsc{Annotation:}\ #2\end{quotation}}
\providecommand{\eprint}[2][]{\url{#2}}

\bibitem{palla2}
Palla G, Barabasi AL, Vicsek T (2007) Quantifying social group evolution.
\newblock Nature 446: 664--667.
\bibAnnoteFile{palla2}

\bibitem{palla1}
Palla G, Derenyi I, Farkas I, Vicsek T (2005) Uncovering the overlapping
  community structure of complex networks in nature and society.
\newblock Nature 435: 814--818.
\bibAnnoteFile{palla1}

\end{thebibliography}

\section*{Figure Legends}

\end{document}


\title{S1: Appendix}
\maketitle

\begin{flushleft}
{\Large
\textbf{Mapping the Evolution of Scientific Fields}
}

Mark Herrera$^{1}$,
David C. Roberts$^{2,\ast}$,
Natali Gulbahce $^{3,\ast}$ \\
\bf{1} Department of Physics and Institute for Research in Electronics and Applied Physics, University of
Maryland, College Park, MD, USA \\
\bf{2} Theoretical Division and Center for Nonlinear Studies, Los Alamos
National Laboratory, Los Alamos, USA

\bf{3} Department of Physics and Center for
Complex Networks Research, Northeastern University, Boston, MA, USA

Center for Cancer Systems Biology, Dana Farber Cancer Institute, Boston, MA \\
$\ast$ E-mail: dcr@lanl.gov, natali.gulbahce@gmail.com
\end{flushleft}

\section{Community Dynamics}
Once community structure is established, a variety of different measurements are performed on the dynamics of the evolving communities.

\begin{figure}[h!] 
\begin{center}
\scalebox{1}{\includegraphics{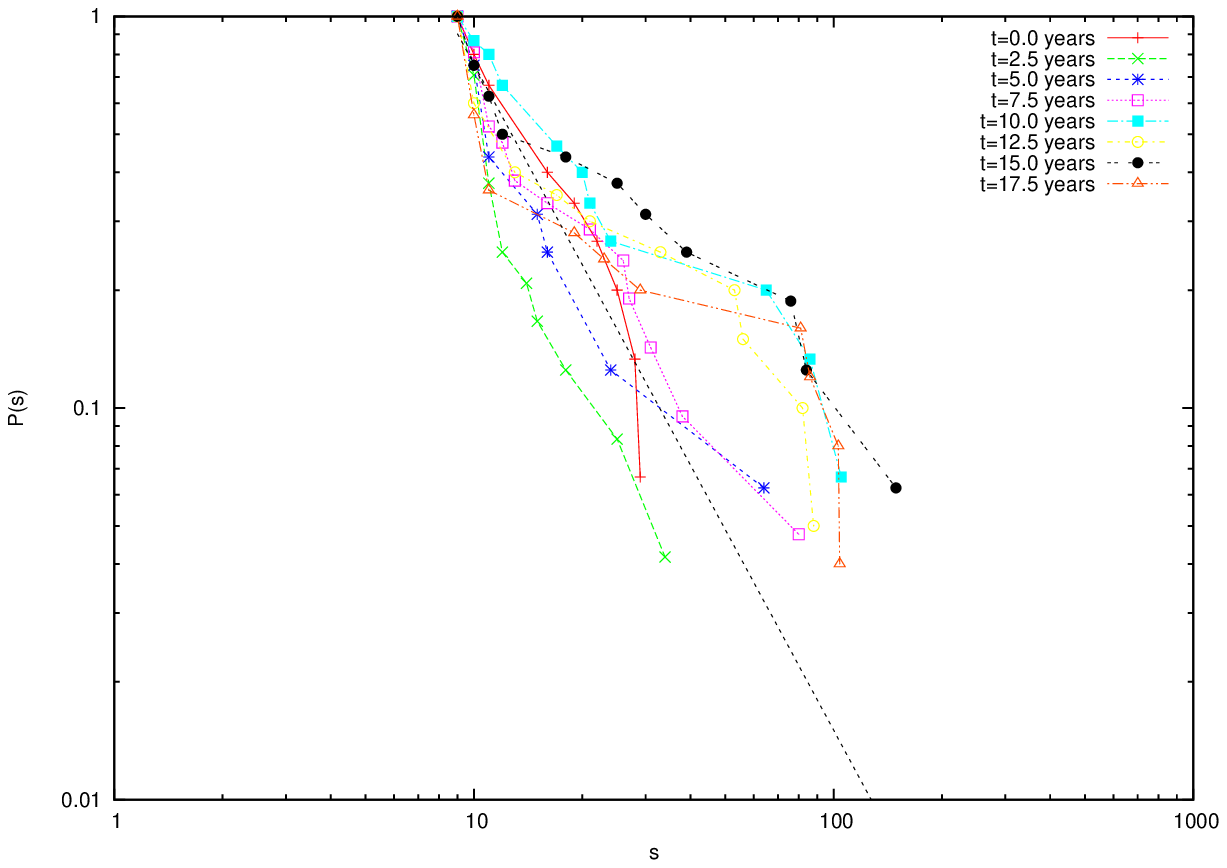}} 
\caption{\label {cumulsize} The
cumulative size distribution for various times in the network.  The
distributions appear long tailed over one decade.} 
\end{center} 
\end{figure}

The cumulative community size distribution appears long tailed over one decade, which is robust as a function of $t$ (years), as shown in Fig. S\ref{cumulsize}.

\begin{figure}[h!] 
\begin{center}
\scalebox{1}{\includegraphics{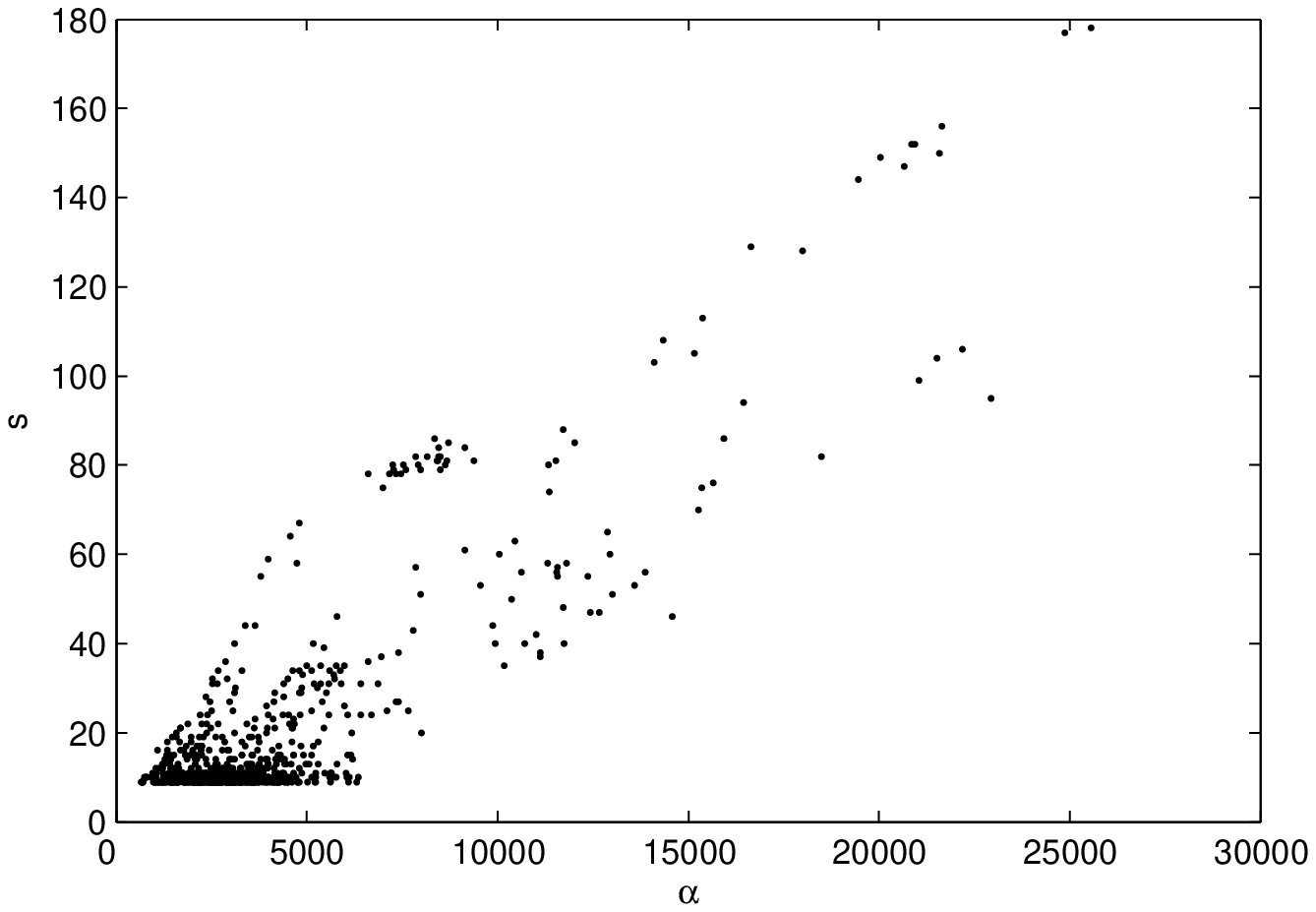}} 
\caption{\label {svsa} {\bf The activity $\alpha$ of each community plotted against its
size $s$ for every time interval ($l=2.5$).}  Notice the positive correlation
between $\alpha$ and $s$. } 
\end{center}
\end{figure}

Fig. S\ref{svsa} plots for all time intervals the size of every
community against its activity $\alpha$ for paper lifetime $l=2.5$.  There appears to be a positive correlation between the two measures, and this trend is observed for $l=0$ (not shown).

\begin{figure}[h!] 
\begin{center}
\scalebox{1}{\includegraphics{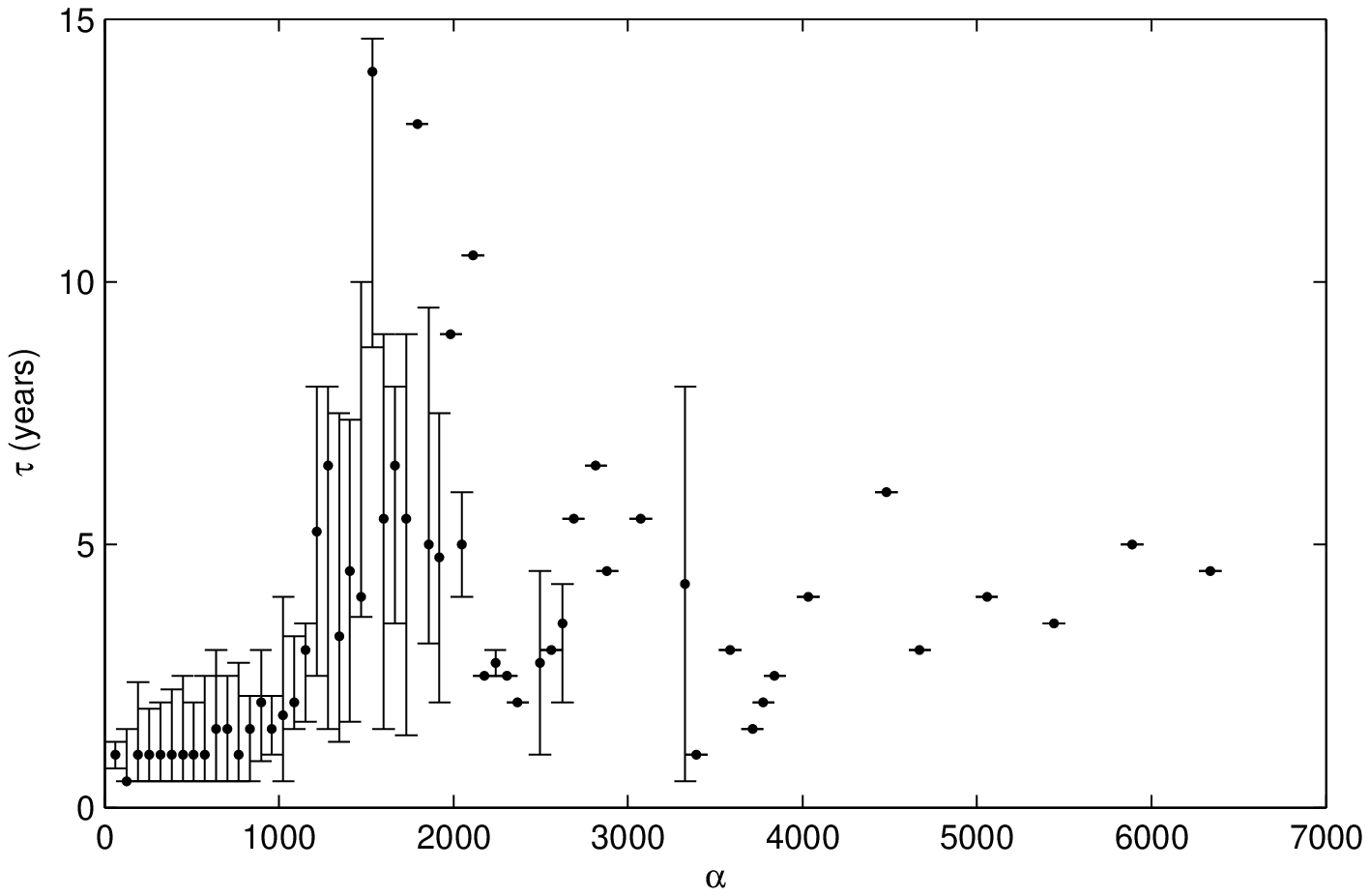}} 
\caption{\label {tauaval0} {\bf The median lifetime as a function of activity for $k=7$, $l= 0$.  }
Notice the trend of $\tau$ increasing with activity.} 
\end{center}
\end{figure}

\begin{figure}[h!] 
\begin{center}
\scalebox{1}{\includegraphics{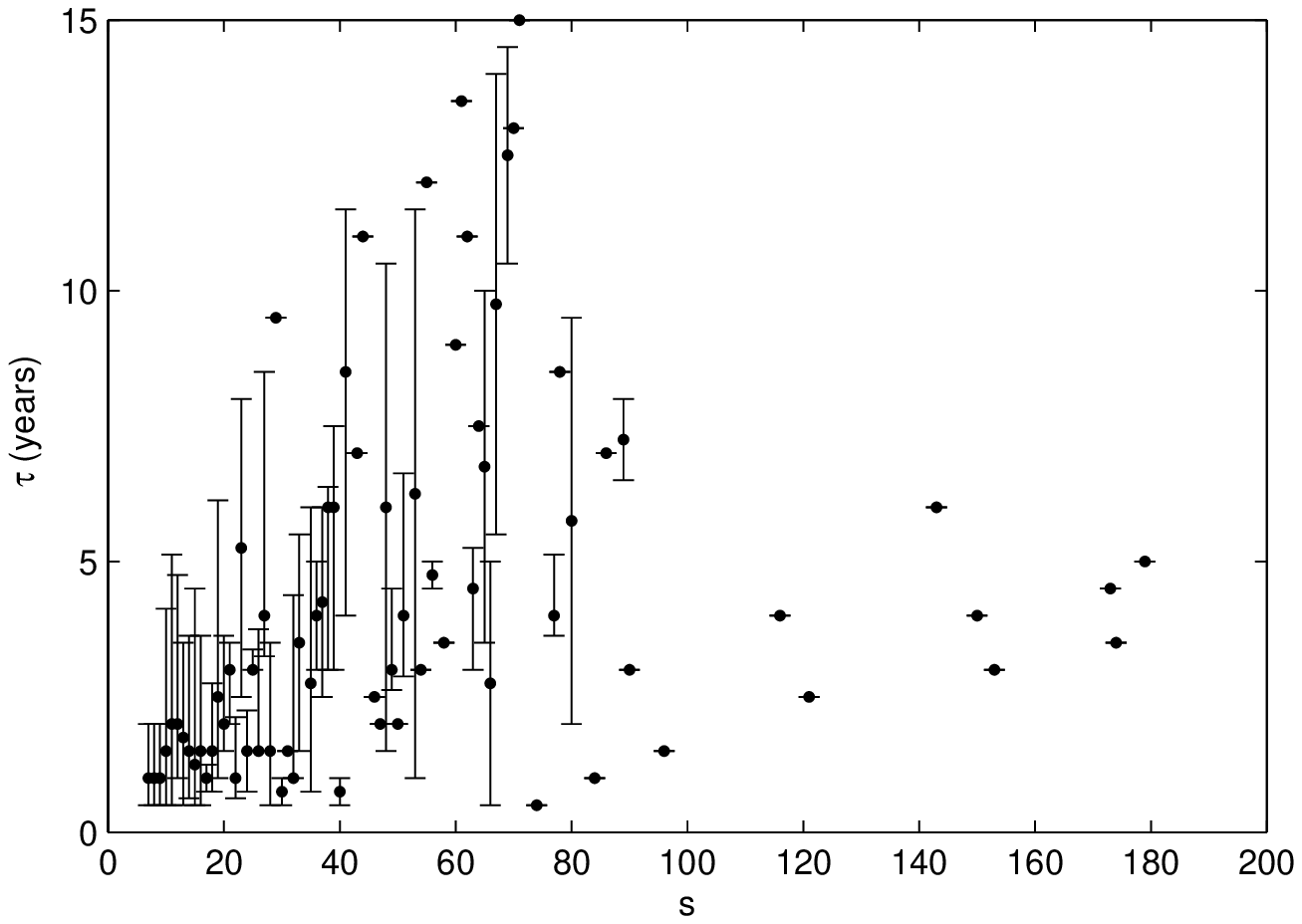}} 
\caption{\label {tausval0} {\bf The median lifetime as a function of size for $k=7$, $l= 0$.  }
Notice the trend of $\tau$ increasing with size.} 
\end{center}
\end{figure}

The dependence of age on activity was measured and Fig.~S\ref{tauaval0} shows the results for the $\tau$
vs. $\alpha$ measurements for $k=7$ with a paper lifetime of $l= 0$ years, where $\alpha$ values were binned because of the wide
range in $\alpha$, as well as to reduce noise.  In both cases  (the one presented here and the one presented in the letter) there is a trend of age increasing as activity increases (though less apparently for $l=0$) and, as expected given the correlation between $\alpha$ and $s$, one sees a similar relationship between age and size, as shown in Fig. S\ref{tausval0}.  Thus, older communities also tend to encompass more publications, a result that
agrees with naive expectation. Further, we note an apparent  phase transition in
 both paper lifetime cases (more apparent for l=2.5 than for $l=0$); after some critical $\alpha$, communities tend to be
longer lived.

Further understanding of the community dynamics can be gained by studying the
volatility of the evolving communities, a measure of how much communities tend
to change between subsequent time steps.  To see this, we define an \emph{age dependent}
running stationarity $\xi(\tau)$ based on community correlation and stationarity
presented by Palla et al.~\cite{palla2}. The correlation $C(t,t')$ between two
states of the same community $A(t)$ at times $t$ and $t'$ is 
\begin{equation*} 
C(t,t')=\left|\frac{A(t) \cap A(t')}{A(t) \cup A(t')}\right|.
\end{equation*} 

\noindent Then the running stationarity, $\xi(\tau)$, of that community is the
average correlation between subsequent time steps up to age $\tau$, 
\begin{equation*}
\xi(\tau)=\frac{1}{\tau-1}\sum_{t'=t_0}^{t_0+\tau-2}C(t',t'+1). 
\end{equation*}

\begin{figure}[h!] 
\begin{center} 
\scalebox{1}{\includegraphics{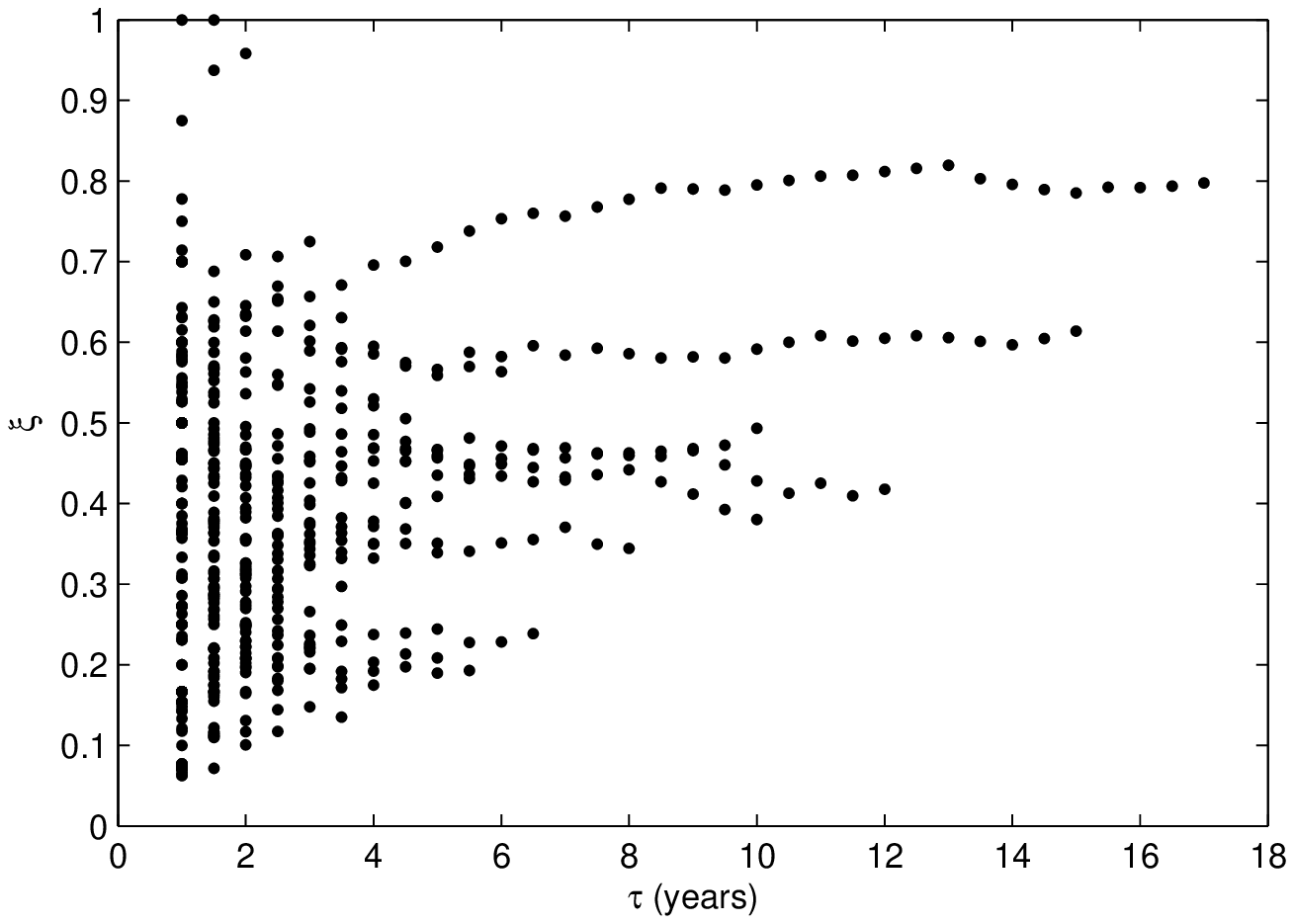}}
\caption{\label {run_statl0} {\bf Age of each community ($k=7$,  $l=0$) vs its running
stationarity value for all time bins.} } 
\end{center} 
\end{figure}

The running stationarity, $\xi(t)$, is plotted against lifetime for every
community with $\tau>1$ along with its current age at time $t$, for all $t$,
for $l=0$ in Fig. S\ref{run_statl0}. This result is qualitatively similar to results obtained using
randomized correlations.  For larger values of $l$, the distribution shifts to
larger values of $\xi(t)$.

\section{Picking a $k$ value}

\begin{figure}[h]
\begin{center}
\scalebox{1}{\includegraphics{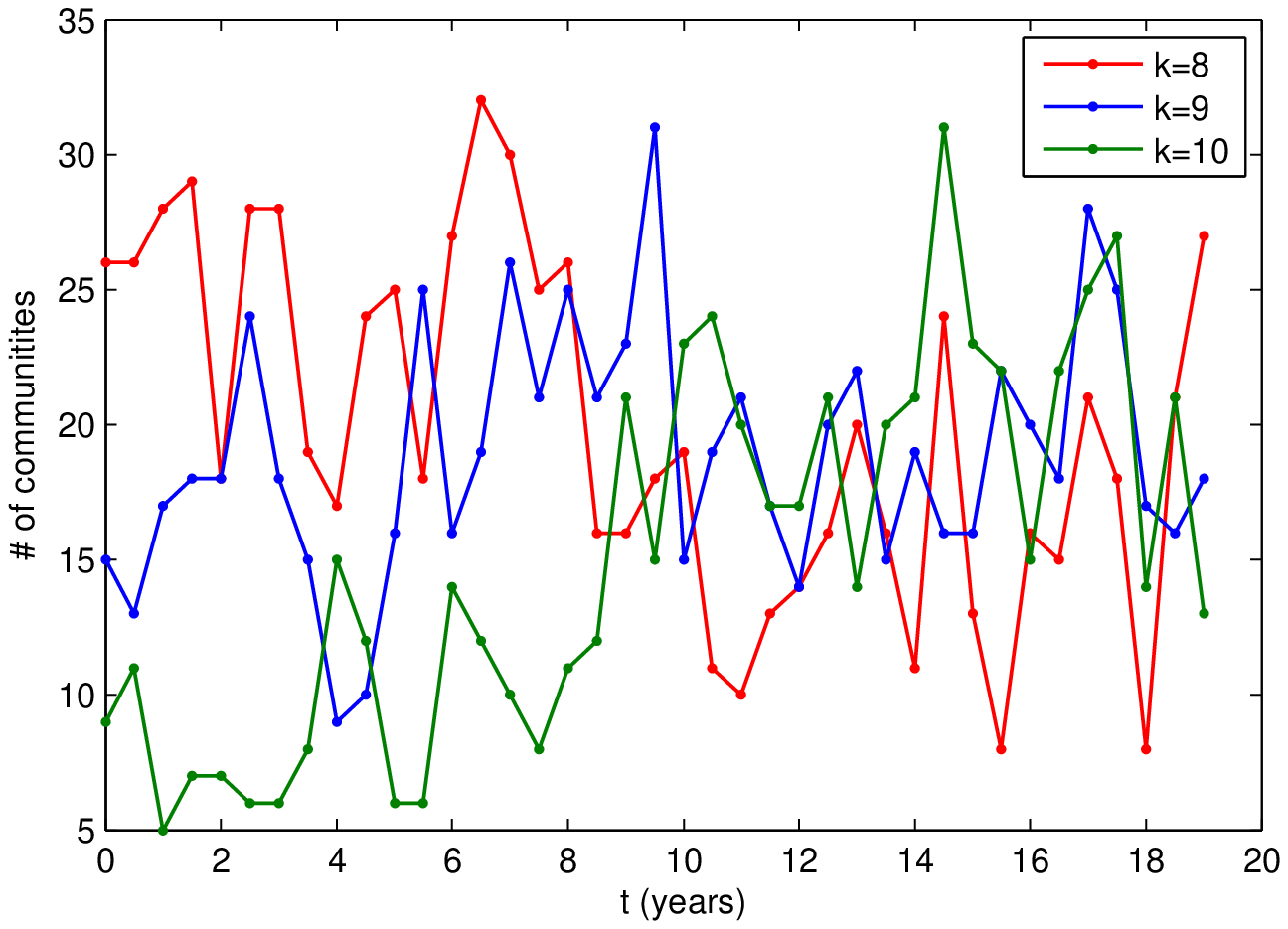}}
\caption{\label {commnum} {\bf The number of communities present in the network
(after the noise measures have been applied) as a function of time for various
$k$ values, with $l=2.5$}. In order to improve the statistical quality of the analysis, larger
numbers of communities are favorable, making $k=10$ an unfavorable parameter
choice. }
\end{center}
\end{figure}

Throughout the paper, $k=9$ is principally used (for l=2.5) because it appears to produce a
large number of communities while discouraging the formation of giant
communities. Further, by keeping $k$ constant, we keep the resolution constant
for the entire analysis. Picking an appropriate $k$ value for the analysis is
done by considering two properties: the number of communities present, and the
presence of overly large communities \cite{palla1}. It is desirable to have a
large number of communities, so as to increase the statistical quality of
measurements made on the network. Fig. S\ref{commnum} plots the number of
present communities for each time step for  $k=8, 9$, and $10$, for $l=2.5$. As
demonstrated, the number of communities found using the choice of $k=10$ tends
to be less than the other parameter choices, making it less favorable in terms
of improving statistical quality.

\begin{figure}[h]
\begin{center}
\scalebox{1}{\includegraphics{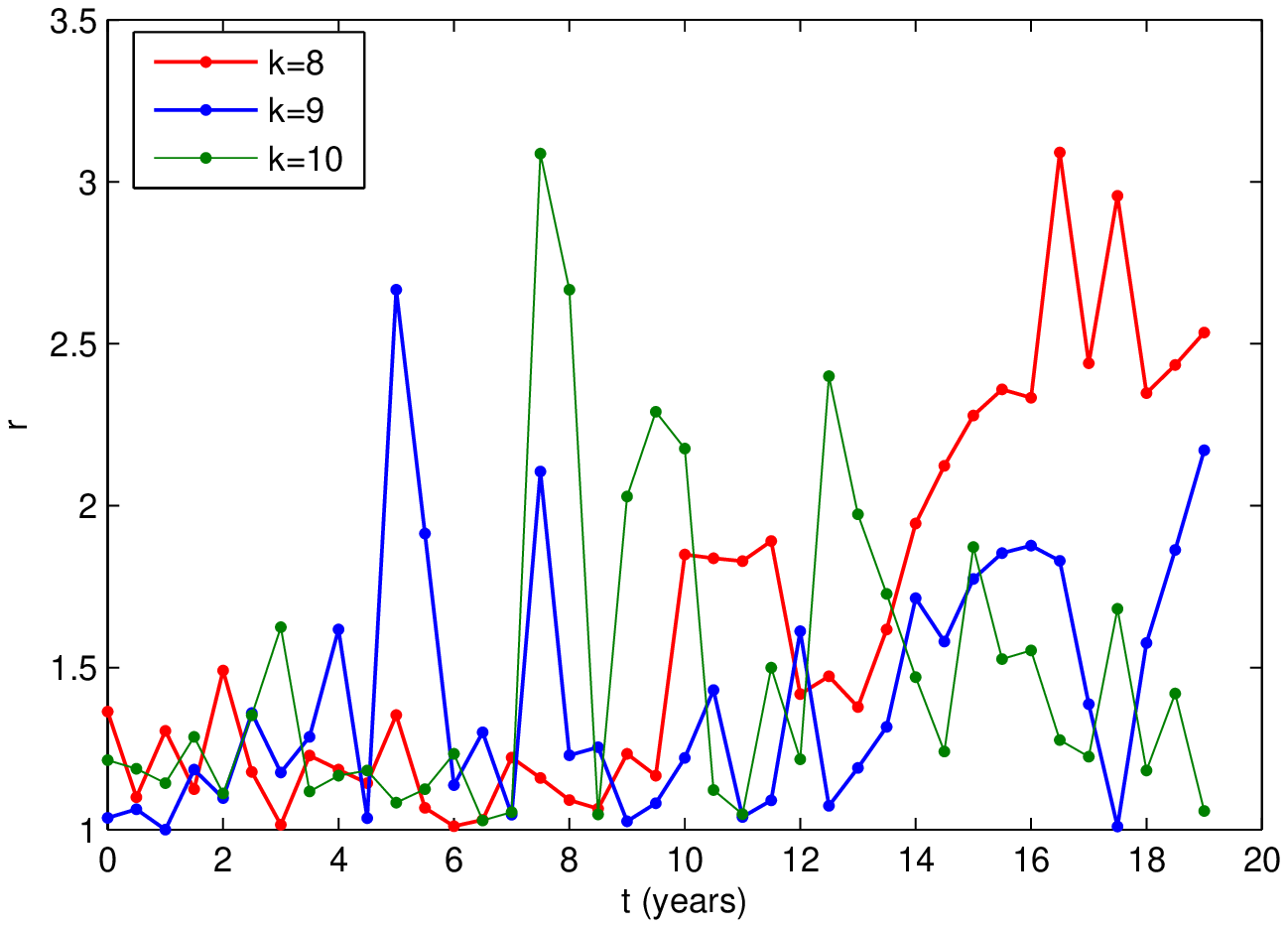}}
\caption{\label {rfig} {\bf The ratio $r$ of the size of the largest community present
divided by size of the second largest community for every time bin for $l=2.5$}. Large $r$
indicates the presence of overly large communities that obscure the community
structure; thus $k=8$ is an unfavorable choice of parameter.}
\end{center}
\end{figure}

A $k$ value must also be large enough to avoid the introduction of overly large
communities that obscure the actual community structure of the network
\cite{palla1}. To quantify this property, we use the quantity $r$ which is the
ratio of the size of the largest community to the second largest community for a
given time bin.  Thus while some distribution in the sizes of communities is
necessary, $r$ should not be overly large.   Fig. S\ref{rfig} plots the measure
$r$ against all time bins for $l=2.5$.  For $k=8$, the values of $r$ tend to become larger (signifying giant communities) than those calculated
from the other two parameter values, making it an unfavorable parameter choice. \\

\section{Merging of communities}
\begin{figure}[h!] 
\begin{center} 
\scalebox{1}{\includegraphics{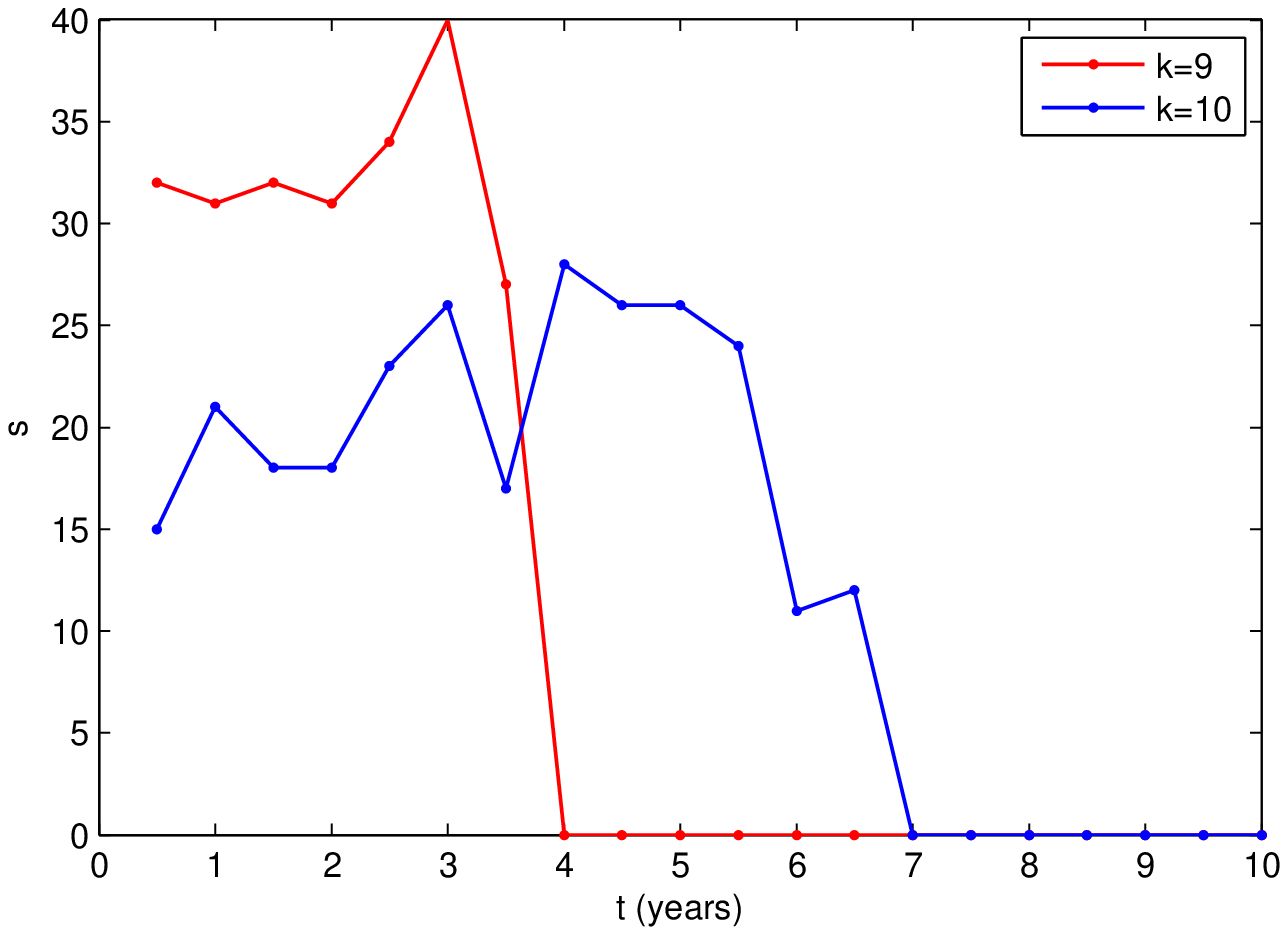}} 
\caption{\label {ctrack} {\bf Size of the nuclear physics community vs time for $k=9$
and $k=10$,  using $l=2.5$.} While the community
appears to die at  $t=8$ ($4$ years) for $k=9$, a community of similar nodes is seen to
continue beyond the time of apparent death when using the higher community
cohesiveness requirement of $k=10$.  It is possible then that the nuclear  physics
community is still present in the analysis, but has merged with another
community. } 
\end{center}
\end{figure}

Lastly, we present an example of a merger between two communities.  Tracking
a nuclear physics community, Fig. S\ref {ctrack} shows the size of
that community as a function of time for $k=9$ and a community of similar nodes with $k=10$, using $l=2.5$.

With $k=9$, it appears that this particle physics community abruptly dies at
$t=4$ years.  Increasing the cohesiveness of communities by increasing to $k=10$
demonstrates that a community composed of similar nodes continues to propagate
past this time of apparent death. Thus, it seems that the nuclear physics community is still present in the network, but has become absorbed by another
community.  

\begin{figure}[h!] 
\begin{center} 
\scalebox{1}{\includegraphics{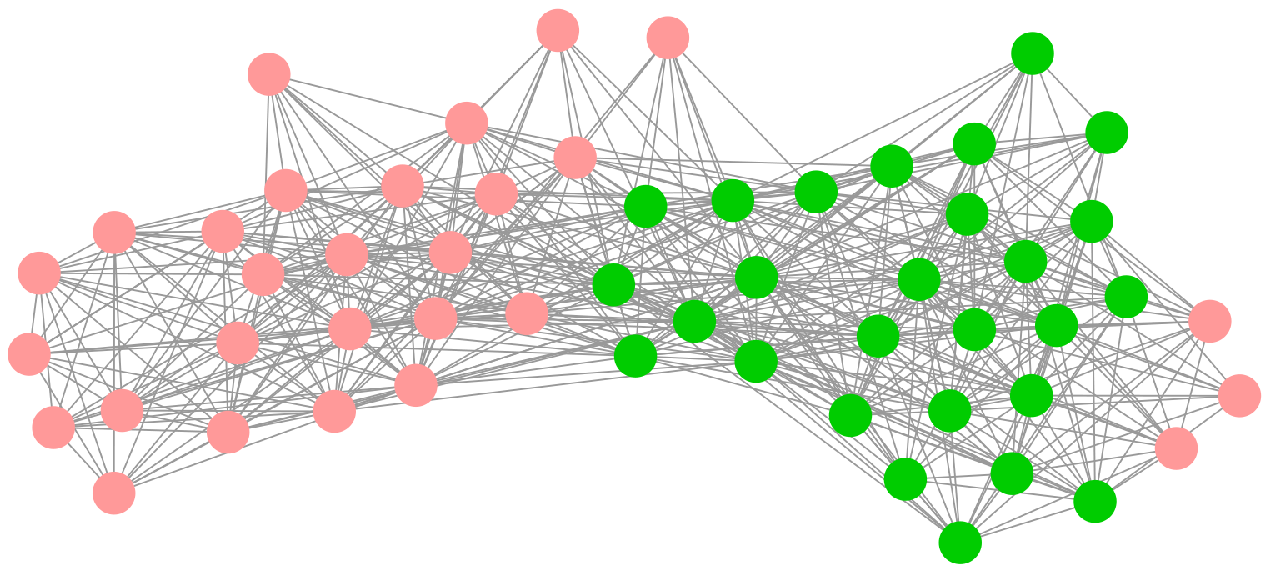}}
\caption{\label {merge} {\bf Merger of the nuclear physics community  (green) with
another community (particle physics: specific reactions and phenomenology) at the time of apparent
death, $t=8$ ($4$ years) for the nuclear physics community}.}
\end{center} 
\end{figure}

Fig. S\ref{merge} plots a community at $t=4$ years with the nodes from
the nuclear physics community displayed in green.  We can assign a label
to this community in the usual manner using the nodes present just before the
apparent death of the nuclear physics community. Doing so, the absorbing
community is comprised of the `physics of elementary particles and fields: specific reactions and phenomenology' in the time bin prior to its absorption of the particle physics community.

\bibliography{plos_new}

\section*{Figure Legends}

